\documentclass[aps,physrev,preprint,groupedaddress]{revtex4-2}

\usepackage{graphicx}
\usepackage[dvipsnames]{xcolor}
\usepackage{siunitx}
\usepackage{amssymb}
\usepackage{amsmath}
\begin{document}


\title{\textbf{A homodyne detection scheme for all-optical photon-photon scattering experiments using 2D detectors}}


\author{Timo Pohle}
\email[]{timo.pohle@physik.lmu.de}
\affiliation{Ludwig-Maximilians-Universität München, Am Coulombwall 1, 85748 Garching, Germany}
\author{Leonard Doyle}
\affiliation{Ludwig-Maximilians-Universität München, Am Coulombwall 1, 85748 Garching, Germany}
\author{Paul Ihde}
\affiliation{Helmholtz Institute Jena, Fröbelstieg 3, 07743 Jena, Germany}
\affiliation{Faculty of Physics and Astronomy, Friedrich-Schiller-Universität Jena, 07743 Jena, Germany}
\author{Felix Karbstein}
\affiliation{Helmholtz Institute Jena, Fröbelstieg 3, 07743 Jena, Germany}
\affiliation{Faculty of Physics and Astronomy, Friedrich-Schiller-Universität Jena, 07743 Jena, Germany}
\author{Jörg Schreiber}
\affiliation{Ludwig-Maximilians-Universität München, Am Coulombwall 1, 85748 Garching, Germany}
\author{Matt Zepf}
\affiliation{Helmholtz Institute Jena, Fröbelstieg 3, 07743 Jena, Germany}
\affiliation{Faculty of Physics and Astronomy, Friedrich-Schiller-Universität Jena, 07743 Jena, Germany}



\date{\today}

\begin{abstract}
Low signal-to-noise ratios are a common problem in experiments attempting to measure photon-photon scattering. In the optical regime, where petawatt lasers with femtosecond pulse durations are used, the large beam sizes cause the major contribution of the background to be spread over up to \SI{100}{\pico\second} in arrival time, whereas the signal is confined to the femtosecond scale. We present a balanced homodyne measurement scheme, which exploits this property to suppress the background. By interfering the signal with a short reference pulse, the measurement becomes effectively gated to the pulse duration and is therefore only sensitive to the co-timed part of the light, reducing the effective background by 3-4 orders of magnitude. Additionally, increasing the reference pulse energy increases the amplitude of the measured quantity without changing the intrinsic signal-to-noise ratio. Using this property, other external noise sources can be made negligible by boosting the measured quantity above the noise floor. Using two-dimensional detectors further enhances the scheme by improving sensitivity and enabling self-referenced single-pulse measurements. In addition, an evaluation procedure based on maximum-likelihood estimation is presented and demonstrated. The robustness and performance of this scheme are demonstrated on simulated data, where a more than 100-fold reduction of measurement time compared to conventional photon-counting methods under realistic conditions is found. 
\end{abstract}


\maketitle

\section{Introduction \label{sec:Introduction}}
The vacuum described by quantum electrodynamics (QED) is not completely empty, but contains virtual electron-positron pairs. This induces an effective interaction between electromagnetic fields, so that light can scatter off light even in vacuum, violating the superposition principle. This becomes a sizable effect in the strong field regime when the field strength approaches the critical field strength of $E_c = 1.3 \cdot 10^{18}\,$\si{\volt\per\meter} set by QED parameters  \cite{Euler1935,Heisenberg1936} and shows up as a redistribution of photons in angle, frequency, or polarization \cite{Karplus,PhysRev.129.2354}.\\
Although this prediction is more than 90 years old, it is still a challenge to test. So far, optical searches have only been able to set upper limits on elastic photon-photon scattering \cite{Moulin1996,Bernard2000,PVLAS,BMV}. Likewise, experiments in the X-ray regime have reported only bounds rather than a signal \cite{INADA2014356,YAMAJI2016454, WATT2025139247}. At higher energies, photon-photon scattering has been reported in ultraperipheral heavy-ion collisions at the LHC by ATLAS and CMS \cite{ATLASLbL,CMS2019}. What is still missing is a direct observation with real photons in the low energy regime, where the probing photons have energies $\hbar\omega \ll m_ec^2$ with $m_e$ being the electron mass. \\
Experiments exploring this regime are currently gaining traction due to the increased availability of petawatt-class lasers and the buildup of multi-petawatt systems like at  NSF OPAL \cite{OPAL}, ELI \cite{ELI_general,ELI_interferometer}, and SEL \cite{Shen_2018}. Using such infrastructure, many collaborations explore photon-photon scattering effects both at large facilities, e.g., BIREF@HIBEF at the European XFEL \cite{birefathibef}, and in smaller experiments like DeLLight at LASERIX \cite{Dellight} and the efforts to measure photon-photon scattering at CALA \cite{Schuetze2024}.\\
In geometries put forward in theoretical proposals, three or more beams, different wavelengths, or asymmetric collision schemes are often used because this causes the signal photons to have a different wavelength, direction, or polarization, which facilitates their separation from the background \cite{Bu2026,different_properties_1,different_properties_2,different_properties_3,different_properties_4,different_properties_5,different_properties_6,Lundin2006,Gies2018}.  However, experiments often opt for simpler geometries because alignment, stability, synchronization, and available laser infrastructure quickly become a limiting factor. In these cases, the signal photons are predominantly elastically scattered and therefore just change their angle slightly in the interaction region. They therefore remain close to the forward cone of the initial probe beam constituting the background from which the scattered photons need to be discriminated; see, e.g., \cite{angle_change_1,angle_change_2,angle_change_3,angle_change_4}.\\
For such a scenario, a strong background suppression is important. With this aim, some experiments, including the one at CALA, employ the dark-field approach \cite{Schuetze2024,jena_dark_field}. This means that an obstacle is introduced into the probe beam before the interaction and the detector is placed into the shadow after it \cite{dark_field_1,dark_field_2,dark_field_3}. Exploiting that scattered photons can fall into the shadow while background photons do not under ideal conditions, this strategy reduces the background by many orders of magnitude. However, aside from direct probe light, other sources of background need to be considered as well. For CALA-like conditions, background studies show that scattering from imperfect optics is the dominant source, while rest-gas scattering is much weaker \cite{Doyle2022Background}. The former, and therefore the majority of the background, is spread over up to \SI{100}{\pico\second} in arrival time because different light paths are taken, while the latter is expected to be nearly co-timed with the vacuum signal ($\mathcal{O}($\SI{10}{\femto\second}$)$). A high temporal selectivity is therefore expected to suppress the background by several orders of magnitude.\\
In this paper, we study an interferometric measurement scheme, closely related to optical homodyne detection \cite{OHD_main,OHD_ex_3,OHD_ex_2,OHD_ex_4, OHD_ex_1, OHD_array1,OHD_array2}, for this exact situation. In Sec.~\ref{sec:homodyne_detection} the basic working principle, the advantage of using 2D detectors, as well as the evaluation strategy, are explained. Then, the performance of the method on simulated data is demonstrated in Sec.~\ref{sec:demo_sim_data}. The results and their implications for the experiment at CALA are discussed in Sec.~\ref{sec:discussion}, before the paper is concluded in Sec.~\ref{sec:conclusion}.

\section{Homodyne Detection as an Optically Gated Measurement \label{sec:homodyne_detection}}
\subsection{Working principle \label{subsec:working_principle}}

\begin{figure}
\includegraphics[width=0.7\columnwidth]{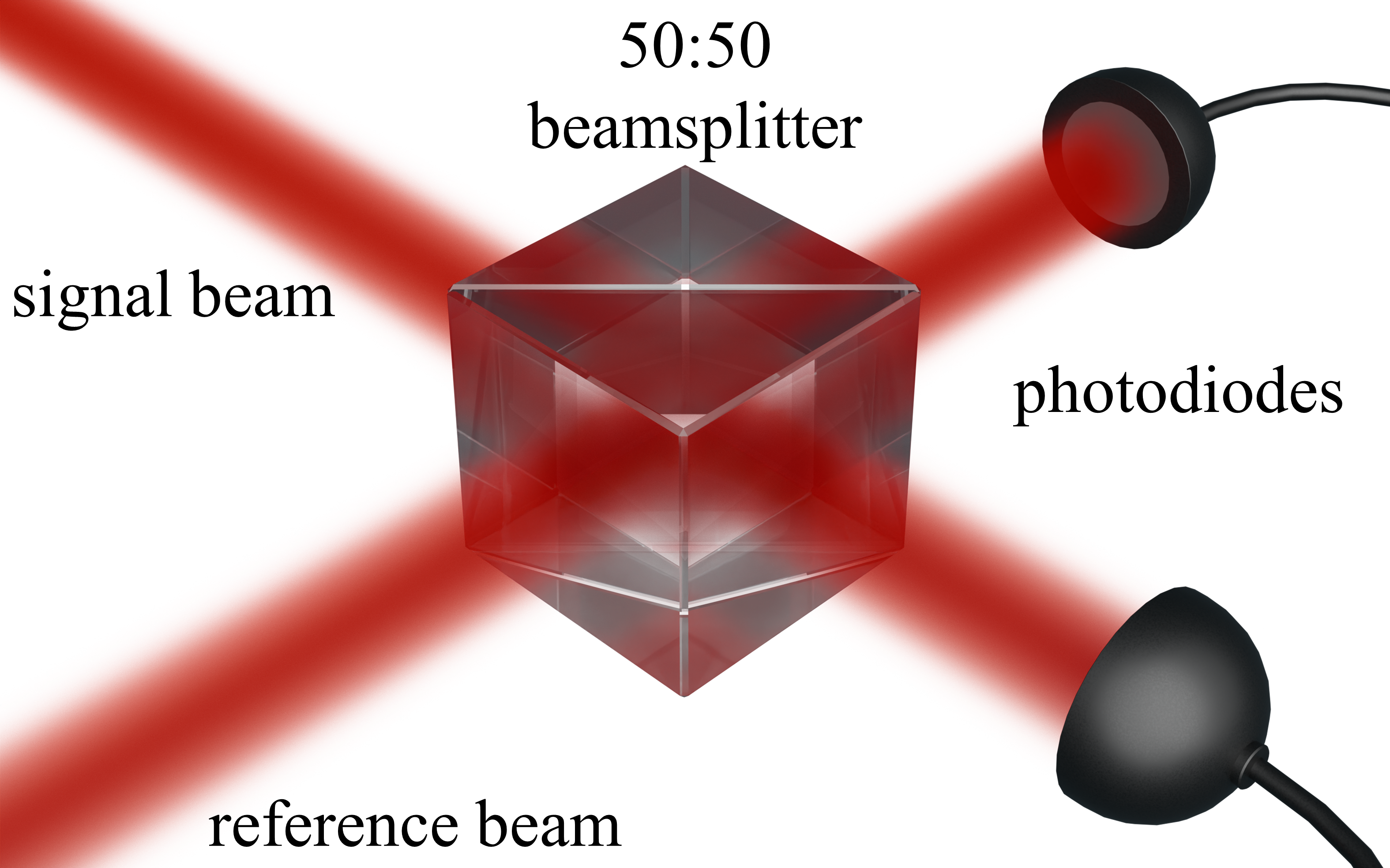}%
\caption{A sketch of the minimal setup required for (balanced) homodyne detection. The interference of a signal and a reference beam is measured by photodiodes simultaneously at the two outputs of a \num{50}:\num{50} beamsplitter.\label{fig:setup}}
\end{figure}

Homodyne detection is an interferometric approach in which the pulse of interest, the signal, is interfered with a well-known reference pulse, the local oscillator (LO), using a \num{50}:\num{50} beamsplitter. For now, the measurement of the two output pulses is assumed to be performed by temporally and spatially integrating detectors, which can be, for example, photodiodes. They are effectively temporally integrating, because the pulse duration and other temporal distributions appearing in the context of this work are much shorter than the bandwidth of available electronics. A sketch of this setup can be seen in Fig.~\ref{fig:setup}.\\
Assuming mean photon numbers $\gamma_{\text{LO}}$ and $\gamma_{\text{sig}}$ in the two pulses and a temporal Gaussian envelope of the electric field of the pulse with a $\frac{1}{e}$-width of $\sigma_t$, the detectors each measure on average a photon number of 
\begin{equation}
     \gamma_{\text{tot},\pm}(\Delta\varphi) = \gamma_{\text{LO}} + \gamma_{\text{sig}} \pm 2\sqrt{\gamma_{\text{LO}}\gamma_{\text{sig}}}\cos{(\Delta\varphi)}\cdot e^{-\frac{\tau^2}{2\sigma_t^2}}.
     \label{eq:interferometer_with_delay}
\end{equation}
Here, $\Delta\varphi$ is the relative optical phase of the two pulses, $\tau$ is their relative delay, and each detector measures the third term with the sign denoted in the subscript. The first two terms are the incoherent contributions by both pulses, while the last term is a cross-term arising due to the coherence of the two pulses. It should be noted that it oscillates with the relative phase $\Delta\varphi$ between the signal and the reference pulse and scales with the square roots of the photon numbers of both pulses. The relative phase $\Delta\varphi$ can be changed in principle in two ways. One option is to change the phase of the electric field relative to its envelope (i.e. the carrier envelope phase) for one of the pulses, which can introduce phase differences $\Delta_{\rm CEP}$ between them. The other, often much more practical method, is to introduce small delays $\tau$ between the two pulses, causing slightly different parts of the pulses to interfere. Of course, this delay must be smaller than the pulse lengths, which is represented by the Gaussian factor in Eq.~\eqref{eq:interferometer_with_delay} multiplying the third term and reducing the contrast of the oscillation for non-zero delays. For pulses with central angular frequency $\omega$, the relative optical phase is therefore described by $\Delta\varphi = \omega\tau + \Delta_{\rm CEP}$.\\
From Eq.~\eqref{eq:interferometer_with_delay}, one can see that the quantity of interest, the signal photon number $\gamma_{\text{sig}}$, can be accessed not only directly by its appearance in the second term, but also indirectly by measuring the amplitude of the oscillation when scanning through $\Delta\varphi$ and knowing the photon number of the local oscillator $\gamma_{\text{LO}}$. Under certain circumstances, such a measurement can be more practical than measuring the signal amplitude directly. As with lock-in amplifiers, this is particularly advantageous, when the signal is hidden in a large, non-coherent noise background from a different source, which dominates the measurement. Such a background manifests as an additional term in Eq.~\eqref{eq:interferometer_with_delay}. By increasing the strength of the local oscillator, the oscillatory term can be boosted to a level above this noise floor, making the measurement possible. At the same time, the uncorrelated noise is not amplified. This is possible for cases where the noise is not intrinsic to the signal and therefore does not contribute to the cross-term. Examples of such backgrounds are high electronic noise in the detector \cite{Homodyne_det_NIR}, or additional incoherent light captured by the detector, which causes additional shot noise.
 The latter kind of background is expected to pose one of the main challenges for all-optical photon-photon scattering experiments. In this context, the background photons fall into two categories. One is the light of the probe pulse being scattered by the electrons of remaining gas atoms in the collision volume, which will be denoted $\gamma_{\text{gas}}$ in the following. The other one is the light of the probe pulse, which scatters off optics and other parts of the experimental setup in such a way that it hits the otherwise shielded detector. As this does not depend on the pressure of the rest gas, it will be denoted as $\gamma_{\text{static}}$. The total background can therefore be written as $\gamma_{\text{bg, tot}} = \gamma_{\text{static}} + \gamma_{\text{gas}}$.
 For the setup at CALA \cite{Schuetze2024}, the static scattering is expected to contribute between $10^4$ and $10^5$ photons, and the rest gas only on the order of 10 photons, which could even be reduced by improving the vacuum chamber. However, the photons contributing to $\gamma_{\text{static}}$ originate from light paths containing at least two scattering events, which causes them to arrive with a much broader temporal distribution at the detector compared to the original probe pulse. For the beam size used in the experiment at CALA, a path length difference of a few centimeters is easily possible, yielding a temporal spread of $10$ to $100$ ps. This is much longer than the photon-photon scattering signal, which is slightly shorter than the probe pulse (\SI{30}{\femto\second}). The background caused by the rest gas, on the other hand, has a temporal distribution like the probe pulse and therefore also similar to the photon-photon scattering signal.
 By using a correctly timed, short pulse as the reference, one can exploit this difference in the temporal distribution of the background contributions. In this case, the whole photon-photon scattering signal interferes with the reference pulse, but only a tiny fraction of the static scattering is present during this short time. This makes the measurement time-gated to the pulse length and the non-co-timed light a source of noise that is not intrinsic to the measurement. In turn, the previously mentioned strategy of increasing the strength of the local oscillator pulse can be used to selectively boost the oscillation term over this background. In the case of the experiment at CALA, the reference pulse length is \SI{30}{\femto\second}. This gating time is much shorter than what is usually possible with electronically time-gated measurement devices, which often have gating times not less than \SI{100}{\pico\second}. This short gating effectively reduces the background by \num{3} to \num{4} orders of magnitude:
  \begin{equation}
     \frac{\gamma_{\text{static,co-timed}}}{\gamma_{\text{static,tot}}} \propto \frac{\sigma_{t}}{\sigma_{t,\text{bg}}}\approx 10^{-3}\dots 10^{-4},
     \label{eq:background_reduction}
 \end{equation}
 where $\sigma_{t}$ is the pulse length and $\sigma_{t,\text{bg}}$ is the temporal width of the background distribution. The co-timed part of the static scattering can therefore be estimated to be $\mathcal{O}(10)$ photons, while the rest of the $10^4$ to $10^5$ photons do not contribute to the cross-term. The remaining, co-timed background consists of a strongly reduced static contribution of $\mathcal{O}(10)$ photons and another $\mathcal{O}(10)$ photons from scattering off the rest gas. The expected signal due to photon-photon scattering, which amounts to approximately \num{3} or less signal photons per pulse, needs to be measured on top of this. In order to include the non-co-timed background, Eq.~\eqref{eq:interferometer_with_delay} can be adapted to
 \begin{equation}
     \gamma_{\text{tot},\pm}(\Delta\varphi) = \gamma_{\text{LO}} + \gamma_{\text{sig}} \pm 2\sqrt{\gamma_{\text{LO}}\gamma_{\text{sig}}}\cos{(\Delta\varphi)}\cdot e^{-\frac{\tau^2}{2\sigma_t^2}} + \gamma_{\text{bg}},
     \label{eq:interferometer_with_bg}
 \end{equation}
  where $\gamma_{\text{bg}}$ is the non-co-timed part of the static background, and $\gamma_{\text{sig}} = \gamma_{\text{static, co-timed}} + \gamma_{\text{gas}} + \gamma_{\text{pps}}$ is all the light in the signal arm that is co-timed with the reference pulse and therefore contributes to the interference pattern. Here, $\gamma_{\text{pps}}$ denotes the photons due to photon-photon scattering, which should be quantified in the experiment.\\
The strategy of increasing the strength of the local oscillator to suppress the incoherent background works due to the scalings of the terms in Eq.~\eqref{eq:interferometer_with_bg}. For high photon numbers in the local oscillator pulse, the intrinsic shot noise caused by the quantized nature of light is given by $\sqrt{\gamma_{\text{tot},\pm}} \approx\sqrt{\gamma_{\text{LO}}}$, because the contribution of the local oscillator dominates. It is important to note that the amplitude of the cross-term, the quantity of interest, scales identically, which results in the signal-to-noise ratio for the oscillation amplitude to be independent of the strength of the local oscillator:
 \begin{equation}
     \text{SNR} \propto \frac{2\sqrt{\gamma_{\text{LO}}\gamma_{\text{sig}}}\cdot e^{-\frac{\tau^2}{2\sigma_t^2}}}{\sqrt{\gamma_{\text{LO}}}} = 2\sqrt{\gamma_{\text{sig}}}\cdot e^{-\frac{\tau^2}{2\sigma_t^2}}.
     \label{eq:SNR}
 \end{equation}
This allows the photon number in the local oscillator pulse to be increased as much as necessary to overcome the noise caused by the incoherent background. The only limit for the photon number in the local oscillator is determined by the dynamic range of the detector, which needs to be able to resolve the oscillatory term growing with $\sqrt{\gamma_{\text{LO}}}$ on top of the constant contribution of the local oscillator, scaling linearly in $\gamma_{\text{LO}}$, i.e., a relative modulation of size $\frac{1}{\sqrt{\gamma_{\text{LO}}}}$.
We can therefore measure a faint, short signal hidden in a large, temporally spread background by interfering the signal with a well-known, short reference pulse and scanning their relative phase to measure the amplitude of the oscillation.\\
Until this point, conducting the measurement is possible with only one of the detectors, but requires precise knowledge of the energy of the reference pulse as well as keeping it constant while scanning the relative phase. The higher the reference energy is chosen, the more accuracy is required because of the previously mentioned scaling. This requirement of a high accuracy can be relaxed by introducing a balanced measurement, which is a common strategy for measurements using optical homodyne detection. For this, both of the outgoing pulses of the beamsplitter, each following Eq.~\eqref{eq:interferometer_with_bg}, are measured and subtracted from each other. Assuming the beamsplitter splits both pulses equally, both detectors measure the same, but with a different sign of the oscillatory term. Therefore, all constant terms cancel in the subtraction, while the oscillation term remains with doubled magnitude:
\begin{equation}
     \Delta\gamma_{\text{tot}} = \gamma_{\text{tot},+} - \gamma_{\text{tot},-}  = 4\sqrt{\gamma_{\text{LO}}\gamma_{\text{sig}}}\cos{(\Delta\varphi)} \cdot e^{-\frac{\tau^2}{2\sigma_t^2}}.
     \label{eq:interferometer_balanced}
 \end{equation}
Note especially that the constant contributions $\gamma_{\text{LO}}$ and $\gamma_{\text{bg}}$ cancel, while noise contributions like the shot noise ($\approx \sqrt{\gamma_{\text{LO}}}$) do not because of their statistical nature. The strength of the reference beam then only needs to be known to extract $\gamma_{\text{sig}}$ from the oscillation amplitude scaling with $\sqrt{\gamma_{\text{LO}}\gamma_{\text{sig}}}$, but not to determine this amplitude. This facilitates the measurement drastically in the case of strong reference beams as only the relative error of the reference beam energy linearly influences the error of the extracted signal strength.

\subsection{Application to 2D detectors \label{subsec:application_to_2D}}
The previously described measurement technique can be further improved by using spatially resolving detectors, which allow for a scan of the relative phase $\Delta\varphi$ in a single-pulse measurement. By introducing a small angle between the reference beam and the signal beam, their optical path lengths to a given point on the detector differ, introducing a relative delay $\tau$ between the pulses. For different positions on the detector, the delay varies, which means different relative phases $\Delta\varphi$. In this scheme, the comparison of different points within one image replaces the previously necessary scanning of $\Delta\varphi$ over many measurements.
This makes the measurement self-referencing using a single pulse, because the full amplitude of the oscillation can be extracted from a single measurement. This relaxes the requirement of constant pulse energy over multiple pulses while scanning $\Delta\varphi$ and eliminates the need for a constant and controllable phase $\Delta\varphi$. Note that each measurement type yields the same amount of information in total regardless of the geometry. The signal-to-noise ratio of each measurement performed by the different pixels on the spatially resolving detectors is lower, as each pixel only measures a fraction of the signal. However, combining all these measurements yields the same overall signal-to-noise ratio as for the single measurement performed by a spatially integrating detector.
Having a self-referenced measurement is important for experiments like the one put forward at CALA, where the stability and control over the phase $\Delta\varphi$ over multiple measurements cannot be guaranteed because the parts of the setup inside and outside of the vacuum chamber may not vibrate equally. In the case of a spatially integrating detector and parallel beams, this would make a controlled scan in $\Delta\varphi$ troublesome. For the case of a 2D detector and a slight angle of the beams, the relative vibrations only shift the fringe pattern, which can easily be taken care of in the evaluation routine as we are only interested in the modulation amplitude.\\
Another advantage of using a 2D detector is the increase in dynamic range. While photodiodes are often read out using analog-to-digital converters (ADC) with a dynamic range of \num{16} bit, a modern complementary metal-oxide-semiconductor (CMOS) sensor can have a higher dynamic range just by the large amount of pixels it contains. Modern camera sensors can also be read out with \num{16} bit, while consisting of millions of pixels, each being almost single-photon-sensitive. This increases the maximum number of photons by a factor of $10^6$ while staying in principle single-photon-sensitive, which increases the dynamic range of the sensor as a whole by the same factor. This holds as long as no spatial information is required and all individual measurements are condensed into a single number, the signal photon count in the present case. A larger dynamic range allows for a stronger reference pulse leading to a better suppression of the background.
The implementation of a balanced detection scheme using two camera sensors is unconventional, because the subtraction is usually performed before digitizing, which is not possible when using off-the-shelf CMOS cameras as the detectors. However, in modern scientific cameras, single photons have a quantifiable effect on the image even after digitizing, which allows for the subtraction to be performed during the evaluation. Capturing the whole signal of both sensors also comes with the benefit that a reliable measurement of the reference beam is available for each pulse by summing both images, which yields $2(\gamma_{\text{LO}} + \gamma_{\text{sig}}+\gamma_{\text{bg}}) \approx 2\gamma_{\text{LO}}$. By using a balanced detection scheme, any structure in the beam profile is detected equivalently by both detectors and cancels when taking their difference. This implies that in combination with the knowledge of the beam profile for each measurement, the method does not pose any strict requirements on the quality of the beam profile.
\subsection{Evaluation \label{subsec:eval_theory}}
To extract the strength of the signal from the measured images, a maximum-likelihood \cite{MLE_first_paper} approach is used because it is a well-established method to reliably determine parameter values and their confidence intervals. It uses a likelihood function based on a model of the experiment that can yield the expected measurements and their distributions for all possible parameter combinations. It is used to judge how likely it is to obtain the measured data for a given set of parameters. The maximum-likelihood estimate of the set of parameters is the one most likely to produce the measured dataset. The model can be found by describing the physical processes at work including their intrinsic uncertainties as well as the measurement process and additional noise sources.\\
Because this is a theoretical study, the model used here is created based on the planned beam properties and the noise sources that are either intrinsic to the process, or that are known from the datasheet of the detector. The model assumes both the reference and the signal beams to be Gaussian. Such a well-behaved distribution would not be necessary as described previously, but is a good approximation for this case because the reference beam can be spatially filtered and the signal originates from a small volume with a size below the diffraction limit of the collecting optic. These two beams are brought to interference yielding an expected value for the photon number at each pixel. In order to describe the uncertainty of this value, all sources of noise need to be included. First, the shot noise of the light on each pixel is modeled using a Poisson distribution, then the quantum efficiency of the camera chip takes action in the form of a binomial distribution. Finally, the additional Poisson-distributed electronic readout noise of the camera is added. Following from the thinning and superposition properties of the Poisson distribution, this hierarchical distribution describing the number of photoelectrons for each pixel can be simplified to a single Poisson distribution with the expected value of
\begin{equation}
    N_{e^-} = \gamma\cdot \text{QE} + \bar{n}_{e},
    \label{eq:electron_number}
\end{equation}
where $\gamma$ is the mean photon number for each pixel, $\text{QE}$ is the quantum efficiency of the camera, and $\bar{n}_{e}$ is the mean number of readout electrons per pixel usually specified in the camera datasheet to quantify electronic readout noise. This distribution of the electron number for each pixel is then mapped to image counts using a linear transformation to account for the camera yielding brightness values, not electron or photon numbers. When calculating the difference of the two Poisson distributions, one for the photoelectrons in the pixels of each of the two cameras, a Skellam distribution is found \cite{Skellam_proof}. Therefore, the likelihood function for one pixel is given by 
\begin{equation}
    \mathcal{L}(\gamma_{\text{sig}}|\Delta g) = \text{Skellam}\left(  \frac{\text{fw}}{2^{N_{\text{bit}}}}\cdot \Delta g; \,\, \text{QE}_+\cdot \gamma_+(\gamma_{\text{sig}})+\bar{n}_+,\,\, \text{QE}_-\cdot \gamma_-(\gamma_{\text{sig}})+\bar{n}_-\right)\cdot \frac{\text{fw}}{2^{N_{\text{bit}}}},
    \label{eq:likelihood_function}
\end{equation}
 where $\Delta g$ is the difference in measured counts for the current pixel position, $\text{fw}$ the full-well capacity of one pixel in electrons, $N_{\text{bit}}$ the bit depth of each pixel, and the subscripts $+$ and $-$ label the two detectors. Note that the two appearances of the factor $\frac{\text{fw}}{2^{N_{\text{bit}}}}$ are the manifestation of the linear mapping between pixel value and photoelectrons in the pixel. Due to computational and practical advantages in the numerical implementation, the corresponding Gaussian approximation to this distribution is used. Great care was taken that the error is negligible, which is mainly guaranteed by the large and similar values of $\gamma_+(\gamma_{\text{sig}})$ and $\gamma_-(\gamma_{\text{sig}})$ making the Skellam distribution effectively Gaussian. Additionally, the evaluation demonstrated below operates on the logarithm of $\mathcal{L}$, i.e., the log-likelihood. This simplifies practical aspects without changing the results.
The parameter being extracted from the data is $\gamma_{\text{sig}}$, the photon number of the signal pulse within the duration of the reference pulse. For simplicity, other parameters like the position of the two beams as well as their relative angle are assumed to be known and constant.
The only exception to this is the relative phase of the two beams, which is challenging to stabilize in setups of the size planned for the experiment at CALA. This is mainly due to the interferometer being both partly in vacuum and in air, which might cause dynamic arm length differences on the micrometer scale. It is therefore an advantage if the evaluation algorithm is able to cope with this condition. For the case of maximum-likelihood estimation, this unknown nuisance parameter can be eliminated by integrating the likelihood along the relative phase weighted by the probability of this phase value being realized, which is uniformly distributed without prior knowledge. This leads to a marginal likelihood in $\gamma_{\text{sig}}$. Because the evaluation of the likelihood function in Eq.~\eqref{eq:likelihood_function} is computationally costly due to the large pixel number, the integration along $\Delta\varphi$ is performed in an efficient, yet accurate manner. This is done by exploiting the fact that the log-likelihood inherits a sinusoidal behavior in $\Delta\varphi$ due to the oscillatory nature of the interference pattern and therefore only a few evaluations are necessary to know the likelihoods for all values of $\Delta\varphi$. With this knowledge, the integration can be carried out analytically.
The same strategy can be applied to other parameters if they turn out to fluctuate, which reduces the stability requirements for the experiment. On the other hand, if the phase $\Delta\varphi$ of the oscillatory term is stable enough to not be integrated out, this could even yield more precise information about the process creating the signal in future, more advanced precision experiments.
Finally, using maximum-likelihood estimation to estimate the signal strength has the crucial advantage that it automatically provides reliable confidence intervals.

\section{Demonstration on Simulated Data \label{sec:demo_sim_data}}

\subsection{Generation of synthetic data \label{subsec:data_gen}}
\begin{figure}
\includegraphics[width=0.9\columnwidth]{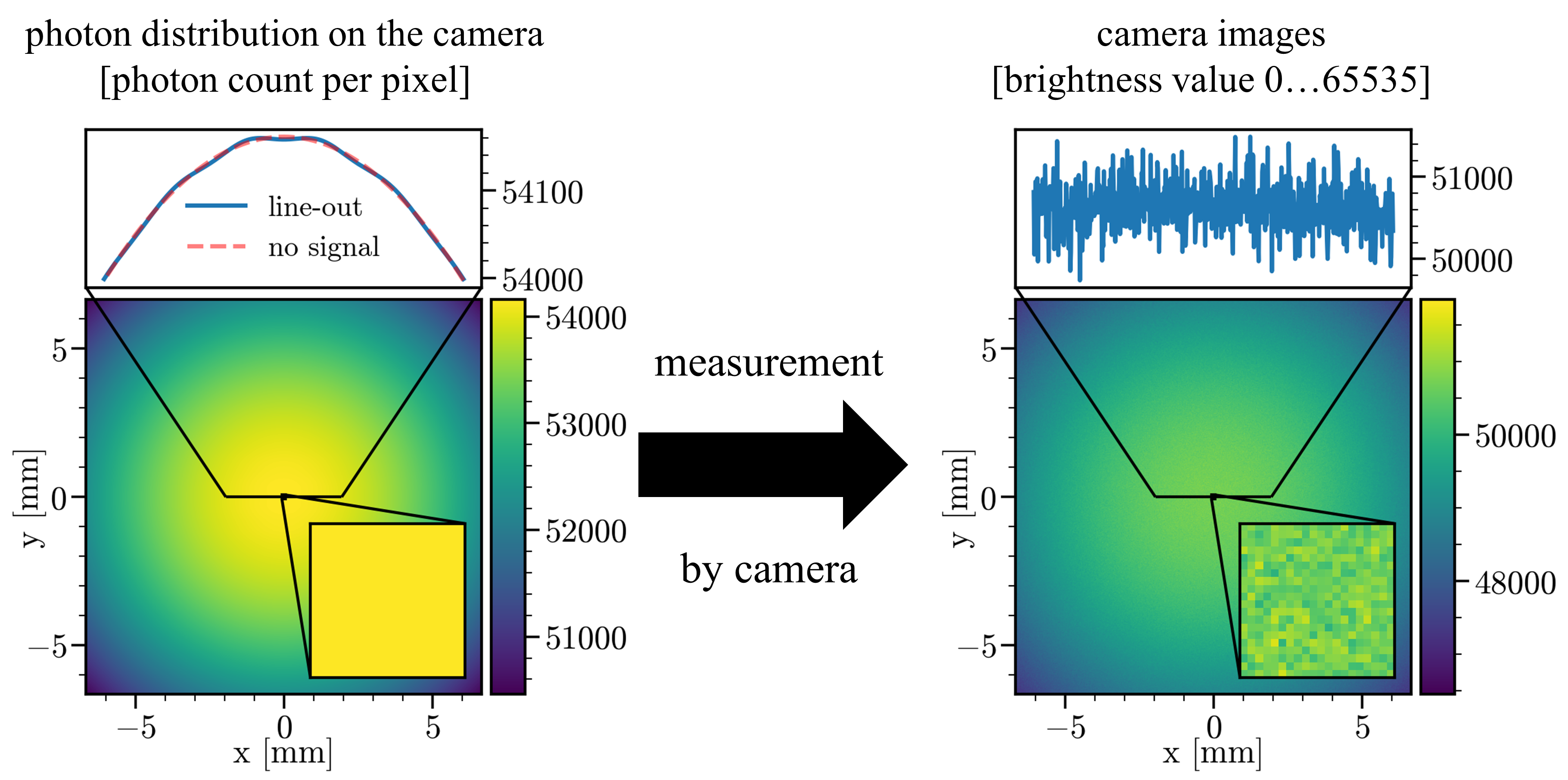}%
\caption{A visualization of how the noise introduced in the measurement process hides the intensity modulations in the photon distribution. The line-out in the left panel shows the small intensity modulations coming from the oscillatory term that needs to be quantified, compared to the case of no modulation when there is no signal pulse.\label{fig:data_gen}}
\end{figure}

To benchmark the measurement strategy and evaluation procedure, we generate synthetic data by calculating the mean photon distribution for a given set of parameters. Noise is introduced by successively sampling from a Poisson distribution to simulate the shot noise, from a binomial distribution to account for the quantum efficiency of the camera chip and then adding a sample from another Poisson distribution to model the electronic noise in the camera. Finally, the data is mapped to the space of pixel brightness and rounded to integers for discretization. This is done for both detectors and for each simulated shot of the laser. This process is repeated to generate a data set containing $N_{\rm S}$ shots.
The properties of the pulses and the detectors were chosen to closely mimic their real counterparts in the experiment at CALA to benchmark the method in realistic conditions. For both cameras, an Andor Marana 4.2B-6 camera is assumed. The readout noise (\num{1.6}$\,e^-$), pixel pitch (\SI{6.5}{\micro\meter}), sensor size (\qtyproduct{13.5 x 13.5}{\milli\meter}), full-well capacity (\num{42000}$\,e^-$) and quantum efficiency ($\approx$ \num{0.6} at \SI{800}{\nano\meter}) stated in the corresponding datasheet \cite{AndorMarana42Bdatasheet} are used.
The LO pulse is chosen to have an energy of \SI{5}{\micro\joule} and a Gaussian beam profile that is centered on the sensor with a \SI{50}{\milli\meter} waist. This is much larger than the sensor and thus ensures a high but nearly homogeneous exposure on a nominal brightness level of around \num{50000} (max. \num{65535} at \num{16} bit).
The signal beam is chosen to be centered on the sensor as well, but with a much smaller waist of \SI{2.5}{\milli\meter} so as not to waste any signal outside of the detector area. The strength of this beam is determined by adding the contributions of the photon-photon scattering signal itself ($\approx$ \num{3} photons per shot \cite{Schuetze2024}), the scattering off the rest gas and the co-timed contribution from the static scattering (each $\mathcal{O}(10)$ photons \cite{Doyle2022Background}). Therefore, \num{42} photons for the total signal pulse is a reasonable choice. The only remaining parameter is the angle between the two beams. It is chosen to be \SI{1}{\milli\radian}, which is high enough for multiple fringes to appear in order to capture the full amplitude of the oscillation while each period is still distributed over tens of pixels to guarantee good sampling.
Figure \ref{fig:data_gen} shows a sketch of the effects of the measurement process. The left panel shows the average photon distribution on one of the cameras for the previously mentioned parameters. The line-out displays the weak intensity modulations on top of the Gaussian beam profile, which are the manifestation of the oscillatory term and have an amplitude of approximately $5$ to $10$ photons per pixel. The right panel shows the same line-out after the measurement by the camera, which introduces noise, most importantly the intrinsic shot noise of the light. The line-out varies so strongly that not even the previously well-visible Gaussian profile of the beam can be recognized anymore. This is in line with expectations because the signal is spread over many pixels resulting in lower signal-to-noise ratios for each individual measurement according to Eq.~\eqref{eq:SNR}. At the same time, it demonstrates the need for an elaborate evaluation routine to extract the information efficiently from the inherently noisy measurement.

\subsection{Evaluation results \label{subsec:eval_results}}
\begin{figure}
\includegraphics[width=0.7\columnwidth]{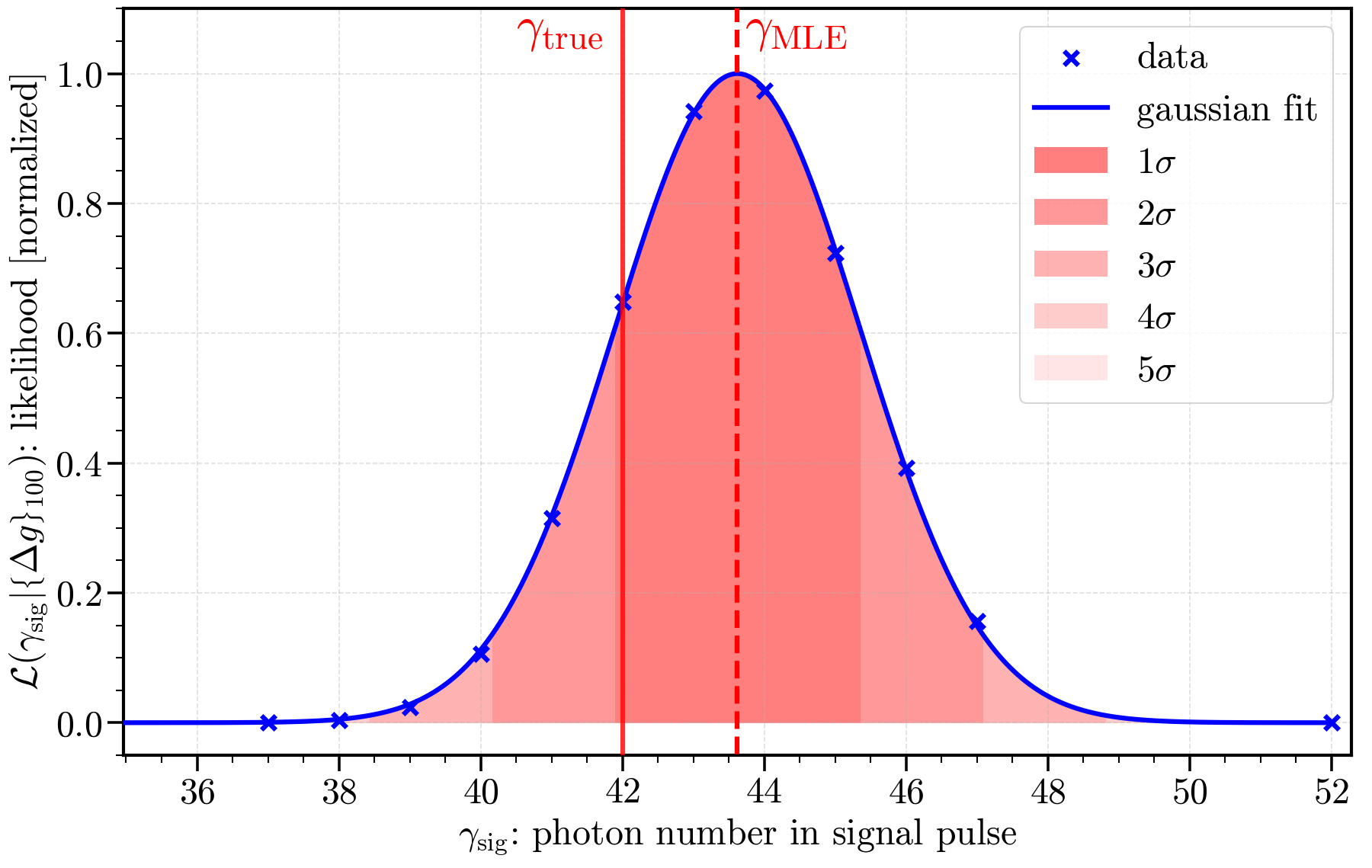}%
\caption{An exemplary likelihood curve resulting from evaluating \num{100} shots of simulated data. The true value for the signal is $\gamma_{\text{true}}=42$, which is within the $1\sigma$ confidence interval around the maximum-likelihood estimate $\gamma_{\text{MLE}}$. \label{fig:example_likelihood_curve}}
\end{figure}

\begin{figure}
\includegraphics[width=0.7\columnwidth]{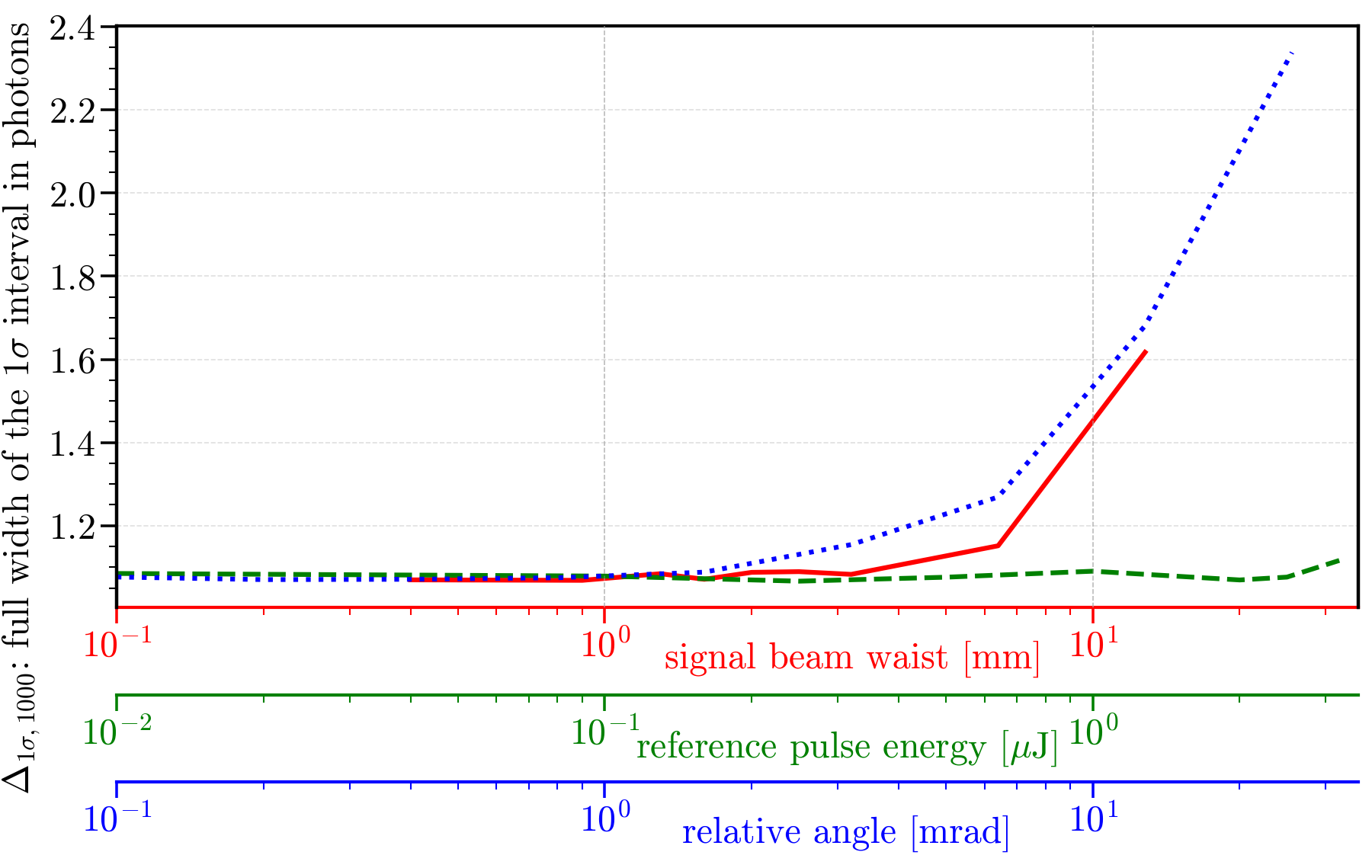}%
\caption{The results of a 1D parameter scan of either the size of the signal beam (solid red), the energy of the reference pulse (dashed green) or their relative angle (dotted blue), while all others are kept at their values described in Sec.~\ref{subsec:data_gen}. The full width of the $1\sigma$ confidence interval extracted from \num{1000} shots ($\Delta_{1\sigma,1000}$) per parameter set is used as a metric for how well the evaluation can extract the information.\label{fig:1D_scans}}
\end{figure}

As described in Sec.~\ref{subsec:eval_theory}, an evaluation employing the maximum-likelihood approach uses a model of the experiment to calculate how likely it is to obtain the measured data for many values of the parameter of interest and then chooses the one with the highest likelihood. In the present case, this parameter is $\gamma_{\rm sig}$ and the analysis is repeated for each synthetic shot. For a given trial value of $\gamma_{\rm sig}$, the model predicts the intensity difference between the two complementary detector images for every pixel. This predicted difference is then compared to the measured pixel-wise difference, and the likelihood defined in Eq.~\eqref{eq:likelihood_function} is evaluated for each pixel. Assuming the pixels are statistically independent, the product of the pixel likelihoods then gives the likelihood of the entire shot for that particular value of $\gamma_{\rm sig}$. Repeating this procedure over many trial values yields a likelihood curve as a function of $\gamma_{\rm sig}$. Its maximum defines the maximum-likelihood estimate (MLE) of $\gamma_{\rm sig}$ for this one shot, while the associated confidence interval is obtained from the shape of the curve. Here, it is extracted by fitting its approximately Gaussian form. Likelihood curves from multiple shots can be combined in the same way, namely by multiplying them to obtain a joint estimate of $\gamma_{\rm sig}$ with smaller confidence intervals.
Figure \ref{fig:example_likelihood_curve} displays an exemplary likelihood curve of a set of \num{100} shots, which was created as described in Sec.~\ref{subsec:data_gen}. Although the likelihood curve was only evaluated for 13 values of $\gamma_{\rm sig}$ due to the high computational cost of each evaluation, the continuous likelihood curve can be well described using the Gaussian fit. For this case, the maximum-likelihood estimate is \num[separate-uncertainty=true]{43.6 +- 1.8} photons in the signal arm, while $\gamma_{\text{sig}}=42$ was used to create the data. The estimate does not coincide with the real input value due to the noisy nature of the data and the limited statistics. However, considering the confidence intervals, the true value is within $1\sigma$ of the prediction, meaning that they are in agreement. This demonstrates that the evaluation can indeed reconstruct the true signal photon number with an accuracy limited by the noise in the data.
\\
The width of the confidence intervals is a measure of the uncertainty of the prediction. Thus, it can be minimized by varying parameters like the size of the signal beam, the strength of the reference beam and their relative angle in order to find the optimal configuration of the setup. The influence of these three parameters is explored in separate parameter scans. The parameter values from Sec.~\ref{subsec:data_gen} are used and only one parameter is varied in each scan. As the figure of merit, the full width of the $1\sigma$ confidence interval of \num{1000}  combined shots ($\Delta_{1\sigma,1000}$) per parameter setting is used. The results of all parameter scans are shown in Fig.~\ref{fig:1D_scans}. The curves display the width $\Delta_{1\sigma,1000}$ while scanning the signal beam waist (red), the reference pulse energy (green), and the relative angle (blue). Many parameter settings yield a relatively constant width $\Delta_{1\sigma,1000}$, which only changes towards extreme parameter values.\\
In the case of the signal waist, the increase of the width $\Delta_{1\sigma,1000}$ for higher values is due to the beam getting large enough for a significant portion of its energy not to end up on the sensor, which obviously reduces the possible signal-to-noise ratio. When the waist size of the signal beam is decreased, the behavior does not change until the beam size becomes smaller than the size of a full oscillation on the chip, which is below the displayed parameter range of Fig.~\ref{fig:1D_scans}. In this case, the self-referencing property of this measurement method is lost, which also renders the exact implementation of the evaluation invalid. The reason is that losing the self-referencing property means a more complex behavior of the log-likelihood when varying the relative phase $\Delta\varphi$, which is not considered in the implementation of the integration along $\Delta\varphi$. This makes it unreliable in this regime, which is why these lower values are not considered. However, the implementation could be changed to include this case with the downside of making the routine computationally more demanding.\\
The scan of the reference pulse energy covers cases from around \num{200} counts per pixel to about \num{64000} counts per pixel. The width $\Delta_{1\sigma,1000}$ stays approximately constant over this range. For energies beyond the data shown in Fig.~\ref{fig:1D_scans}, the sensor saturates, hiding any extractable information. For lower energies, the width $\Delta_{1\sigma,1000}$ is expected to increase when the oscillation term has a similar or smaller amplitude than the noise induced by the static background. Because this background consists of $10^4$ to $10^5$ photons spread over approximately a million pixels, even for the lowest energy setting in this scan, the oscillation term has a much larger amplitude, making the effect of the background negligible.\\
The scan of the relative angle shows that for small values, i.e., larger fringe spacing, there is the same problem of not sampling a whole oscillation as previously for small waists of the signal beam. Consequently, this direction was not explored further. For large angles, the width $\Delta_{1\sigma,1000}$ clearly increases, i.e., the performance decreases, because of two factors. Firstly, one oscillation spans only a few pixels, which can cause mismatches between the model and data because the model used in the likelihood calculation does not integrate over the pixel area but only predicts the value for the pixel center. Secondly, the pulsed nature of both beams determines the coherence length to be shorter than the path length differences at opposite sides of the beam profile. This leads to the interference pattern appearing only in parts of the profile, while the signal in other parts does not contribute.\\
In summary, the parameter scans demonstrate that the presented method performs well for a wide range of parameters, making it well suited for application in the experiment. This insensitivity to the exact choice of beam parameters also means that it is applicable for a broad range of geometries and suggests that the results of the evaluation put forward in this paper can likely be transferred to other setups.

\begin{figure}
\includegraphics[width=0.7\columnwidth]{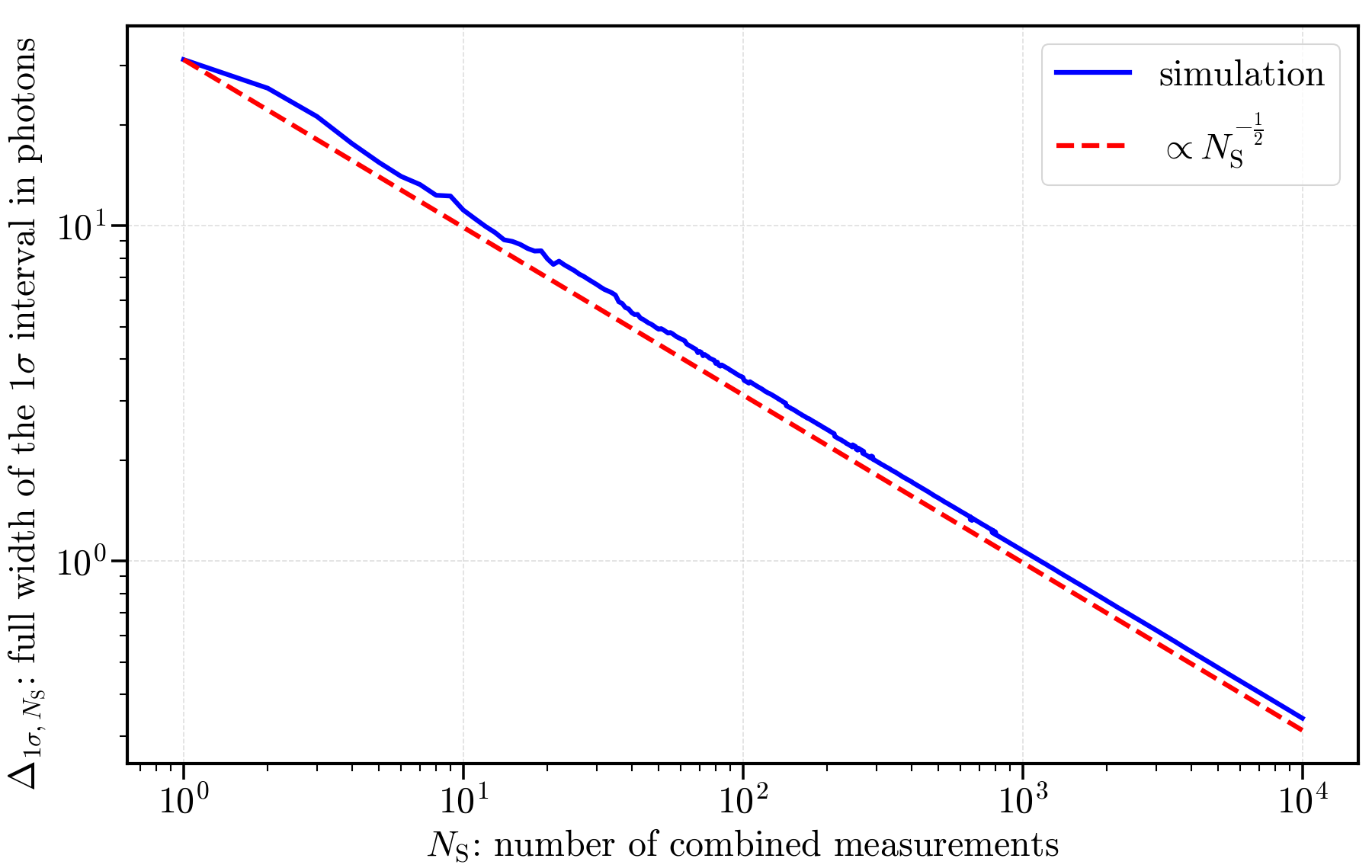}%
\caption{The behavior of the full width of the $1\sigma$ confidence interval when evaluating data sets of different sizes $N_{\rm S}$ in blue (all parameters as described in Sec.~\ref{subsec:data_gen}). For comparison, the expected $\frac{1}{\sqrt{N_{\rm S}}}$ scaling extrapolated from the evaluation of a single shot in red.\label{fig:performance_eval}}
\end{figure}

In order to evaluate the performance of the measurement method and evaluation strategy, a data set consisting of \num{10000} shots is generated for the parameters introduced in Sec.~\ref{subsec:data_gen} and subsequently evaluated along the lines described previously. Subsets of many different sizes $N_{\rm S}$ are chosen and their evaluation results are combined. Again, the full width of the $1\sigma$ confidence interval $\Delta_{1\sigma,N_{\rm S}}$ is examined as a measure of the quality of the prediction for the signal photon yield. Figure \ref{fig:performance_eval} shows its behavior with increasing size of the subset $N_{\rm S}$. The width $\Delta_{1\sigma,N_{\rm S}}$ of a chosen subset scales approximately as $\frac{1}{\sqrt{N_{\rm S}}}$, which is in line with the naive expectation. It can also be derived analytically using the Gaussian form of the model and the additivity of Fisher information while assuming constant measurement conditions \cite{Cowan1998}. While Fig.~\ref{fig:performance_eval} only shows the full width of the $1\sigma$ interval, the full interval of $n\cdot\sigma$ is $n$ times wider as a direct consequence of the Gaussian nature of the likelihood function in use. One can see for example that a full width of $\Delta_{1\sigma,N_{\rm S}} = 6$ photons, meaning a $1\sigma$ error of $\pm 3$ photons, is already reached after around $N_{\rm S} = 36$ shots. Hence, under ideal conditions, it only takes \num{36} seconds of measurement time to
reach a $1\sigma$ statistical uncertainty comparable to the expected photon-photon scattering signal for the parameters used at CALA.
\section{Discussion\label{sec:discussion}}
Because not all photons constituting the signal pulse originate from photon-photon scattering, the background contribution to the signal arm must be quantified separately from the full signal consisting of the background and the photon-photon scattered photons. Taking the difference of these two types of measurement then quantifies the strength of the photon-photon scattering. To detect the expected photon-photon scattering signal with a certain confidence level, the corresponding confidence interval of this difference needs to exclude \num{0}, i.e., the measurement of just the background and the one including the photon-photon scattered photons must differ significantly. Based on the analysis and the assumptions about the signal and background used in the present work, confidence levels of $1\sigma$, $3\sigma$, and $5\sigma$ are reached after at least  approximately \num{140}, \num{1190}, and \num{3222} accumulated shots equally distributed over both types of measurements, respectively. As this assumes an ideal scattering yield and no losses from scattering to detection, this is only a lower bound for the required number of shots. Nonetheless, these numbers imply measurement times below one hour using a laser system with \SI{1}{\hertz} repetition rate, which is well within the range of a realistic experimental campaign.
One can also compare these numbers with the number of required shots if a direct measurement of the photon-photon scattered photons obscured by the large background was performed. In this case, direct measurement means that a time-integrating, photon-counting detector is used. Here, the previous strategy of performing two types of measurements to subtract the background from the full signal is chosen again. Both the background alone and the photon-photon scattered photons together with the background are quantified separately with an equal number of measurements. By taking their difference, the number of photon-photon scattered photons can be found. At CALA, approximately \num{3} photons originating from photon-photon scattering need to be measured against $\gamma_{\text{bg, tot}}$, the combination of both the co- and non-co-timed background. It is assumed to be around \num{75000} photons. Due to the large shot noise of $\sqrt{75\,000}\approx 274$ photons in each of the two types of measurements, a lot of statistics is needed to reliably detect signals on the few-photon level. The confidence intervals after $N_{\rm S}$ accumulated shots can be calculated using the fact that the difference of two Poisson-distributed quantities (the measurements of the two detectors) is Skellam distributed, which can be very well approximated by a Gaussian for the case of both Poisson distributions having a high and similar mean (here $\lambda_1 \approx 75\,000,\, \lambda_2 \approx 75\,003$). The $1\sigma$ confidence interval can be found to be $\sigma = \frac{\sqrt{150\,003}}{\sqrt{N_{\rm S}}}$ after $N_{\rm S}$ accumulated shots. Therefore, at least approximately \num{17000}, \num{150000}, and \num{420000} measurements are needed to verify the existence of the \num{3} scattered photons with a confidence level of $1\sigma$, $3\sigma$, and $5\sigma$, respectively. Hence, using the presented measurement scheme instead of performing a direct photon-counting measurement sizably reduces the number of required shots by a factor of approximately \num{125}.
\section{Conclusion \label{sec:conclusion}}
The homodyne-based detection method presented in this work substantially increases the selectivity of the detector by using a femtosecond pulse as a gate in combination with boosting the oscillatory term beyond the non-co-timed noise. The extension to 2D detectors in conjunction with the maximum-likelihood-based evaluation routine represents a robust way to perform the measurement even in the presence of shot-to-shot fluctuations of parameters such as the relative phase in this work. Benchmarking with simulated data revealed the robustness of the method over a wide range of experimental parameters. Finally, it could be demonstrated that the number of laser shots, and hence measurement time, required to reach a particular confidence level is reduced more than 100-fold compared to a conventional photon-counting measurement. This prospect of reducing the required measurement time from multiple days to below an hour renders experimental measurements feasible in the near future.

\begin{acknowledgments}
We would like to thank Pooyan Khademi and Fabian Schütze for fruitful discussions.
This work has been funded by the Deutsche Forschungsgemeinschaft (DFG) under Grant Nos. 416702141 and  411451547 within the Research Unit FOR2783/2.
\end{acknowledgments}
\bibliography{paper_bibliography}

@PREAMBLE{
 "\providecommand{\noopsort}[1]{}" 
 # "\providecommand{\singleletter}[1]{#1}%" 
}

@article{Euler1935,
author={Euler, H.
and Kockel, B.},
title={{\"U}ber die {Streuung} von {Licht} an {Licht} nach der {Diracschen Theorie}},
journal={Naturwissenschaften},
year={1935},
month={Apr},
day={01},
volume={23},
number={15},
pages={246-247},
issn={1432-1904},
doi={10.1007/BF01493898},
url={https://doi.org/10.1007/BF01493898}
}

@Article{Heisenberg1936,
author={Heisenberg, W.
and Euler, H.},
title={Folgerungen aus der {Diracschen Theorie} des {Positrons}},
journal={Zeitschrift f{\"u}r Physik},
year={1936},
month={Nov},
day={01},
volume={98},
number={11},
pages={714-732},
issn={0044-3328},
doi={10.1007/BF01343663},
url={https://doi.org/10.1007/BF01343663}
}

@article{Karplus,
  title = {The Scattering of Light by Light},
  author = {Karplus, Robert and Neuman, Maurice},
  journal = {Phys. Rev.},
  volume = {83},
  issue = {4},
  pages = {776--784},
  numpages = {0},
  year = {1951},
  month = {Aug},
  publisher = {American Physical Society},
  doi = {10.1103/PhysRev.83.776},
  url = {https://link.aps.org/doi/10.1103/PhysRev.83.776}
}

@Article{Moulin1996,
author={Moulin, F.
and Bernard, D.
and Amiranoff, F.},
title={Photon-photon elastic scattering in the visible domain},
journal={Z. Phys. C},
year={1996},
month={Dec},
day={01},
volume={72},
number={4},
pages={607-611},
issn={1431-5858},
doi={10.1007/s002880050282},
url={https://doi.org/10.1007/s002880050282}
}

@Article{Bernard2000,
author={Bernard, D.
and Moulin, F.
and Amiranoff, F.
and Braun, A.
and Chambaret, J. P.
and Darpentigny, G.
and Grillon, G.
and Ranc, S.
and Perrone, F.},
title={Search for stimulated photon-photon scattering in vacuum},
journal={The European Physical Journal D - Atomic, Molecular, Optical and Plasma Physics},
year={2000},
month={Mar},
day={01},
volume={10},
number={1},
pages={141-145},
issn={1434-6079},
doi={10.1007/s100530050535},
url={https://doi.org/10.1007/s100530050535}
}

@article{INADA2014356,
title = {Search for photon–photon elastic scattering in the X-ray region},
journal = {Phys. Lett. B},
volume = {732},
pages = {356-359},
year = {2014},
issn = {0370-2693},
doi = {https://doi.org/10.1016/j.physletb.2014.03.054},
url = {https://www.sciencedirect.com/science/article/pii/S0370269314002160},
author = {T. Inada and T. Yamaji and S. Adachi and T. Namba and S. Asai and T. Kobayashi and K. Tamasaku and Y. Tanaka and Y. Inubushi and K. Sawada and M. Yabashi and T. Ishikawa}
}

@article{ATLASLbL,
  title = {Observation of Light-by-Light Scattering in Ultraperipheral $\mathrm{Pb}+\mathrm{Pb}$ Collisions with the {ATLAS} Detector},
  author = {{ATLAS Collaboration}},
  journal = {Phys. Rev. Lett.},
  volume = {123},
  issue = {5},
  pages = {052001},
  numpages = {21},
  year = {2019},
  month = {Jul},
  publisher = {American Physical Society},
  doi = {10.1103/PhysRevLett.123.052001},
  url = {https://link.aps.org/doi/10.1103/PhysRevLett.123.052001}
}

@article{CMS2019,
title = {Evidence for light-by-light scattering and searches for axion-like particles in ultraperipheral {PbPb} collisions at {$\sqrt{sNN}=5.02$ TeV}},
journal = {Phys. Lett. B},
volume = {797},
pages = {134826},
year = {2019},
issn = {0370-2693},
doi = {https://doi.org/10.1016/j.physletb.2019.134826},
url = {https://www.sciencedirect.com/science/article/pii/S0370269319305404},
author = {{CMS Collaboration}},
}

@article{Lundin2006,
  title = {Analysis of four-wave mixing of high-power lasers for the detection of elastic photon-photon scattering},
  author = {Lundin, J. and Marklund, M. and Lundstr\"om, E. and Brodin, G. and Collier, J. and Bingham, R. and Mendon\ifmmode \mbox{\c{c}}\else \c{c}\fi{}a, J. T. and Norreys, P.},
  journal = {Phys. Rev. A},
  volume = {74},
  issue = {4},
  pages = {043821},
  numpages = {10},
  year = {2006},
  month = {Oct},
  publisher = {American Physical Society},
  doi = {10.1103/PhysRevA.74.043821},
  url = {https://link.aps.org/doi/10.1103/PhysRevA.74.043821}
}

@article{Gies2018,
  title = {Photon-photon scattering at the high-intensity frontier},
  author = {Gies, Holger and Karbstein, Felix and Kohlf\"urst, Christian and Seegert, Nico},
  journal = {Phys. Rev. D},
  volume = {97},
  issue = {7},
  pages = {076002},
  numpages = {7},
  year = {2018},
  month = {Apr},
  publisher = {American Physical Society},
  doi = {10.1103/PhysRevD.97.076002},
  url = {https://link.aps.org/doi/10.1103/PhysRevD.97.076002}
}

@article{Schuetze2024,
  title = {Dark-field setup for the measurement of light-by-light scattering with high-intensity lasers},
  author = {Sch\"utze, Fabian and Doyle, Leonard and Schreiber, J\"org and Zepf, Matt and Karbstein, Felix},
  journal = {Phys. Rev. D},
  volume = {109},
  issue = {9},
  pages = {096009},
  numpages = {14},
  year = {2024},
  month = {May},
  publisher = {American Physical Society},
  doi = {10.1103/PhysRevD.109.096009},
  url = {https://link.aps.org/doi/10.1103/PhysRevD.109.096009}
}

@article{Doyle2022Background,
  author = {Doyle, L and Khademi, P and Hilz, P and Sävert, A and Schäfer, G and Schreiber, J and Zepf, M},
  title = {Experimental estimates of the photon background in a potential light-by-light scattering study},
  journal = {New J. Phys.},
  volume = {24},
  pages = {025003},
  year = {2022},
  doi = {10.1088/1367-2630/ac4ad3}
}

@Article{birefathibef,
author={Ahmadiniaz, N. and others},
title={Towards a vacuum birefringence experiment at the {Helmholtz} International Beamline for Extreme Fields (Letter of Intent of the {BIREF@HIBEF} Collaboration)},
journal={High Power Laser Sci. Eng.},
year={2025},
edition={2025/03/11},
publisher={Cambridge University Press},
volume={13},
pages={e7},
issn={2095-4719},
doi={10.1017/hpl.2024.70},
url={https://doi.org/10.1017/hpl.2024.70}
}

@article{Dellight,
  title = {Experiment to observe an optically induced change of the vacuum index},
  author = {Robertson, Scott and Mailliet, Aur\'elie and Sarazin, Xavier and Couchot, Fran{\c {c}}ois and Baynard, Elsa and Demailly, Julien and Pittman, Moana and Djannati-Ata\"{\i}, Arache and Kazamias, Sophie and Urban, Marcel},
  journal = {Phys. Rev. A},
  volume = {103},
  issue = {2},
  pages = {023524},
  numpages = {20},
  year = {2021},
  month = {Feb},
  publisher = {American Physical Society},
  doi = {10.1103/PhysRevA.103.023524},
  url = {https://link.aps.org/doi/10.1103/PhysRevA.103.023524}
}

@article{jena_dark_field,
  title = {Proof-of-principle experiment for the dark-field detection concept for measuring vacuum birefringence},
  author = {{\v S}m{\'\i}d, Michal and others},
  journal = {Phys. Rev. A},
  volume = {112},
  issue = {6},
  pages = {063512},
  numpages = {10},
  year = {2025},
  month = {Dec},
  publisher = {American Physical Society},
  doi = {10.1103/xpxy-ntwz},
  url = {https://link.aps.org/doi/10.1103/xpxy-ntwz}
}

@article{
Homodyne_det_NIR,
author = {O. Wolley  and S. Mekhail  and P.-A. Moreau  and T. Gregory  and G. Gibson  and G. Leuchs  and M. J. Padgett },
title = {Near single-photon imaging in the shortwave infrared using homodyne detection},
journal = {Proc. Natl. Acad. Sci. U.S.A.},
volume = {120},
number = {10},
pages = {e2216678120},
year = {2023},
doi = {10.1073/pnas.2216678120}
}

@article{OHD_ex_1,
  title = {Maximum-likelihood estimation of photon-number distribution from homodyne statistics},
  author = {Banaszek, Konrad},
  journal = {Phys. Rev. A},
  volume = {57},
  issue = {6},
  pages = {5013--5015},
  numpages = {0},
  year = {1998},
  month = {Jun},
  publisher = {American Physical Society},
  doi = {10.1103/PhysRevA.57.5013},
  url = {https://link.aps.org/doi/10.1103/PhysRevA.57.5013}
}

@article{OHD_ex_2,
  title = {Photon-number statistics from the phase-averaged quadrature-field distribution: Theory and ultrafast measurement},
  author = {Munroe, M. and Boggavarapu, D. and Anderson, M. E. and Raymer, M. G.},
  journal = {Phys. Rev. A},
  volume = {52},
  issue = {2},
  pages = {R924--R927},
  numpages = {0},
  year = {1995},
  month = {Aug},
  publisher = {American Physical Society},
  doi = {10.1103/PhysRevA.52.R924},
  url = {https://link.aps.org/doi/10.1103/PhysRevA.52.R924}
}

@article{OHD_ex_3,
author = {Bing Qi and Pavel Lougovski and Brian P. Williams},
journal = {Opt. Express},
number = {2},
pages = {2276--2290},
publisher = {Optica Publishing Group},
title = {Characterizing photon number statistics using conjugate optical homodyne detection},
volume = {28},
month = {Jan},
year = {2020},
url = {https://opg.optica.org/oe/abstract.cfm?URI=oe-28-2-2276},
doi = {10.1364/OE.383358},
}

@article{OHD_ex_4,
  title = {Homodyne measurement of the average photon number},
  author = {Webb, J. G. and Ralph, T. C. and Huntington, E. H.},
  journal = {Phys. Rev. A},
  volume = {73},
  issue = {3},
  pages = {033808},
  numpages = {7},
  year = {2006},
  month = {Mar},
  publisher = {American Physical Society},
  doi = {10.1103/PhysRevA.73.033808},
  url = {https://link.aps.org/doi/10.1103/PhysRevA.73.033808}
}

@article{PhysRev.129.2354,
  title = {Nonlinear Interaction of Light in a Vacuum},
  author = {Mckenna, J. and Platzman, P. M.},
  journal = {Phys. Rev.},
  volume = {129},
  issue = {5},
  pages = {2354--2360},
  numpages = {0},
  year = {1963},
  month = {Mar},
  publisher = {American Physical Society},
  doi = {10.1103/PhysRev.129.2354},
  url = {https://link.aps.org/doi/10.1103/PhysRev.129.2354}
}

@article{YAMAJI2016454,
title = {An experiment of X-ray photon–photon elastic scattering with a Laue-case beam collider},
journal = {Phys. Lett. B},
volume = {763},
pages = {454-457},
year = {2016},
issn = {0370-2693},
doi = {https://doi.org/10.1016/j.physletb.2016.11.003},
url = {https://www.sciencedirect.com/science/article/pii/S0370269316306645},
author = {T. Yamaji and T. Inada and T. Yamazaki and T. Namba and S. Asai and T. Kobayashi and K. Tamasaku and Y. Tanaka and Y. Inubushi and K. Sawada and M. Yabashi and T. Ishikawa}
}

@article{WATT2025139247,
title = {Bounding elastic photon-photon scattering at $\sqrt{s} \approx$ 1 {MeV} using a laser-plasma platform},
journal = {Phys. Lett. B},
volume = {861},
pages = {139247},
year = {2025},
issn = {0370-2693},
doi = {https://doi.org/10.1016/j.physletb.2025.139247},
url = {https://www.sciencedirect.com/science/article/pii/S0370269325000073},
author = {R. Watt and others}
}

@article{Shen_2018,
doi = {10.1088/1361-6587/aaa7fb},
url = {https://doi.org/10.1088/1361-6587/aaa7fb},
year = {2018},
month = {feb},
publisher = {IOP Publishing},
volume = {60},
number = {4},
pages = {044002},
author = {Shen, Baifei and Bu, Zhigang and Xu, Jiancai and Xu, Tongjun and Ji, Liangliang and Li, Ruxin and Xu, Zhizhan},
title = {Exploring vacuum birefringence based on a 100 {PW} laser and an x-ray free electron laser beam},
journal = {Plasma Phys. Control. Fusion}
}

@Article{Bu2026,
author={Bu, Zhigang
and Zhang, Lingang
and Liu, Shiyu
and Shen, Baifei
and Li, Ruxin
and Ivanov, Igor P.
and Ji, Liangliang},
title={Super light-by-light scattering in vacuum induced by intense vortex lasers},
journal={Commun. Phys.},
year={2026},
month={Mar},
day={10},
volume={9},
number={1},
pages={144},
issn={2399-3650},
doi={10.1038/s42005-026-02556-0},
url={https://doi.org/10.1038/s42005-026-02556-0}
}

@article{OPAL,
    author = {Rinderknecht, Hans G. and Dill, E. and MacLeod, A. J. and King, B. and Sow, K. and Bahk, S.-W. and Begishev, I. A. and Karbstein, F. and Schreiber, Jörg and Zepf, M. and Di Piazza, A.},
    title = {On measuring stimulated photon–photon scattering using multiple ultraintense lasers},
    journal = {Phys. Plasmas},
    volume = {32},
    number = {8},
    pages = {083301},
    year = {2025},
    month = {08},
    issn = {1070-664X},
    doi = {10.1063/5.0272791},
    url = {https://doi.org/10.1063/5.0272791}
}

@article{different_properties_1,
  title = {Using High-Power Lasers for Detection of Elastic Photon-Photon Scattering},
  author = {Lundstr\"om, E. and Brodin, G. and Lundin, J. and Marklund, M. and Bingham, R. and Collier, J. and Mendon\ifmmode \mbox{\c{c}}\else \c{c}\fi{}a, J. T. and Norreys, P.},
  journal = {Phys. Rev. Lett.},
  volume = {96},
  issue = {8},
  pages = {083602},
  numpages = {4},
  year = {2006},
  month = {Mar},
  publisher = {American Physical Society},
  doi = {10.1103/PhysRevLett.96.083602},
  url = {https://link.aps.org/doi/10.1103/PhysRevLett.96.083602}
}

@article{different_properties_2,
  title = {Orbital Angular Momentum Coupling in Elastic Photon-Photon Scattering},
  author = {Aboushelbaya, R. and Glize, K. and Savin, A. F. and Mayr, M. and Spiers, B. and Wang, R. and Collier, J. and Marklund, M. and Trines, R. M. G. M. and Bingham, R. and Norreys, P. A.},
  journal = {Phys. Rev. Lett.},
  volume = {123},
  issue = {11},
  pages = {113604},
  numpages = {5},
  year = {2019},
  month = {Sep},
  publisher = {American Physical Society},
  doi = {10.1103/PhysRevLett.123.113604},
  url = {https://link.aps.org/doi/10.1103/PhysRevLett.123.113604}
}

@Article{different_properties_3,
author={King, Ben
and Di Piazza, Antonino
and Keitel, Christoph H.},
title={A matterless double slit},
journal={Nature Photonics},
year={2010},
month={Feb},
day={01},
volume={4},
number={2},
pages={92-94},
issn={1749-4893},
doi={10.1038/nphoton.2009.261},
url={https://doi.org/10.1038/nphoton.2009.261}
}

@article{different_properties_4,
  author  = {A. A. Varfolomeev},
  title   = {Induced Scattering of Light by Light},
  journal = {Sov. Phys. JETP},
  volume  = {23},
  number  = {4},
  pages   = {681},
  year    = {1966}
}

@article{different_properties_5,
  title = {Three-pulse photon-photon scattering},
  author = {King, B. and Hu, H. and Shen, B.},
  journal = {Phys. Rev. A},
  volume = {98},
  issue = {2},
  pages = {023817},
  numpages = {10},
  year = {2018},
  month = {Aug},
  publisher = {American Physical Society},
  doi = {10.1103/PhysRevA.98.023817},
  url = {https://link.aps.org/doi/10.1103/PhysRevA.98.023817}
}

@article{different_properties_6,
  title = {Photon merging and splitting in electromagnetic field inhomogeneities},
  author = {Gies, Holger and Karbstein, Felix and Seegert, Nico},
  journal = {Phys. Rev. D},
  volume = {93},
  issue = {8},
  pages = {085034},
  numpages = {13},
  year = {2016},
  month = {Apr},
  publisher = {American Physical Society},
  doi = {10.1103/PhysRevD.93.085034},
  url = {https://link.aps.org/doi/10.1103/PhysRevD.93.085034}
}

@article{dark_field_1,
  title = {Enhancing quantum vacuum signatures with tailored laser beams},
  author = {Karbstein, Felix and Mosman, Elena A.},
  journal = {Phys. Rev. D},
  volume = {101},
  issue = {11},
  pages = {113002},
  numpages = {6},
  year = {2020},
  month = {Jun},
  publisher = {American Physical Society},
  doi = {10.1103/PhysRevD.101.113002},
  url = {https://link.aps.org/doi/10.1103/PhysRevD.101.113002}
}

@article{dark_field_2,
  title = {Direct Accessibility of the Fundamental Constants Governing Light-by-Light Scattering},
  author = {Karbstein, Felix and Ullmann, Daniel and Mosman, Elena A. and Zepf, Matt},
  journal = {Phys. Rev. Lett.},
  volume = {129},
  issue = {6},
  pages = {061802},
  numpages = {6},
  year = {2022},
  month = {Aug},
  publisher = {American Physical Society},
  doi = {10.1103/PhysRevLett.129.061802},
  url = {https://link.aps.org/doi/10.1103/PhysRevLett.129.061802}
}

@article{dark_field_3,
author = {J. Peatross and J. L. Chaloupka and D. D. Meyerhofer},
journal = {Opt. Lett.},
number = {13},
pages = {942--944},
publisher = {Optica Publishing Group},
title = {High-order harmonic generation with an annular laser beam},
volume = {19},
month = {Jul},
year = {1994},
url = {https://opg.optica.org/ol/abstract.cfm?URI=ol-19-13-942},
doi = {10.1364/OL.19.000942},
}

@article{angle_change_1,
doi = {10.1088/1367-2630/ac51a7},
url = {https://doi.org/10.1088/1367-2630/ac51a7},
year = {2022},
month = {mar},
publisher = {IOP Publishing},
volume = {24},
number = {2},
pages = {025010},
author = {Roso, Luis and Lera, Roberto and Ravichandran, Smrithan and Longman, Andrew and He, Calvin Z and Pérez-Hernández, José Antonio and Apiñaniz, Jon I and Smith, Lucas D and Fedosejevs, Robert and Hill, Wendell T},
title = {Towards a direct measurement of the quantum-vacuum Lagrangian coupling coefficients using two counterpropagating super-intense laser pulses},
journal = {New J. Phys.}
}

@article{angle_change_2,
  title = {Vacuum birefringence in strong inhomogeneous electromagnetic fields},
  author = {Karbstein, Felix and Gies, Holger and Reuter, Maria and Zepf, Matt},
  journal = {Phys. Rev. D},
  volume = {92},
  issue = {7},
  pages = {071301},
  numpages = {6},
  year = {2015},
  month = {Oct},
  publisher = {American Physical Society},
  doi = {10.1103/PhysRevD.92.071301},
  url = {https://link.aps.org/doi/10.1103/PhysRevD.92.071301}
}

@article{angle_change_3,
  title = {Light by light diffraction in vacuum},
  author = {Tommasini, Daniele and Michinel, Humberto},
  journal = {Phys. Rev. A},
  volume = {82},
  issue = {1},
  pages = {011803},
  numpages = {4},
  year = {2010},
  month = {Jul},
  publisher = {American Physical Society},
  doi = {10.1103/PhysRevA.82.011803},
  url = {https://link.aps.org/doi/10.1103/PhysRevA.82.011803}
}

@article{OHD_main,
title = {Measuring the quantum state of light},
journal = {Prog. Quantum Electron.},
volume = {19},
number = {2},
pages = {89-130},
year = {1995},
issn = {0079-6727},
doi = {https://doi.org/10.1016/0079-6727(94)00007-L},
url = {https://www.sciencedirect.com/science/article/pii/007967279400007L},
author = {U. Leonhardt and H. Paul}
}

@article{angle_change_4,
  title = {All-optical quantum vacuum signals in two-beam collisions},
  author = {Gies, Holger and Karbstein, Felix and Klar, Leonhard},
  journal = {Phys. Rev. D},
  volume = {106},
  issue = {11},
  pages = {116005},
  numpages = {18},
  year = {2022},
  month = {Dec},
  publisher = {American Physical Society},
  doi = {10.1103/PhysRevD.106.116005},
  url = {https://link.aps.org/doi/10.1103/PhysRevD.106.116005}
}

@article{MLE_first_paper,
    author = {Fisher, R. A.},
    title = {On the mathematical foundations of theoretical statistics},
    journal = {Philos. Trans. R. Soc. Lond. A.},
    volume = {222},
    number = {594-604},
    pages = {309-368},
    year = {1922},
    month = {01},
    issn = {0264-3952},
    doi = {10.1098/rsta.1922.0009}
}

@article{ELI_interferometer,
doi = {10.1088/1402-4896/ade5d7},
url = {https://doi.org/10.1088/1402-4896/ade5d7},
year = {2025},
month = {jun},
publisher = {IOP Publishing},
volume = {100},
number = {7},
pages = {075537},
author = {Ataman, S and Nakamiya, Y},
title = {Vacuum birefringence measurement schemes employing high-power lasers and a {Mach–Zehnder} interferometer},
journal = {Phys. Scr.}
}

@article{ELI_general,
doi = {10.1088/1361-6633/aacfe8},
url = {https://doi.org/10.1088/1361-6633/aacfe8},
year = {2018},
month = {aug},
publisher = {IOP Publishing},
volume = {81},
number = {9},
pages = {094301},
author = {Gales, S and others},
title = {The extreme light infrastructure—nuclear physics ({ELI-NP}) facility: new horizons in physics with 10 {PW} ultra-intense lasers and 20 {MeV} brilliant gamma beams},
journal = {Rep. Prog. Phys.}
}

@article{Skellam_proof,
 ISSN = {09528385, 23972335},
 URL = {http://www.jstor.org/stable/2981372},
 author = {J. G. Skellam},
 journal = {J. R. Stat. Soc.},
 number = {3},
 pages = {296--296},
 publisher = {[Oxford University Press, Royal Statistical Society]},
 title = {The Frequency Distribution of the Difference Between Two {Poisson} Variates Belonging to Different Populations},
 urldate = {2026-05-08},
 volume = {109},
 year = {1946}
}

@book{Cowan1998,
  author    = {Glen Cowan},
  title     = {Statistical Data Analysis},
  publisher = {Oxford University Press},
  year      = {1998}
}

@misc{AndorMarana42Bdatasheet,
  title        = {{Andor Marana sCMOS} specifications},
  author       = {{Andor Technology}},
  howpublished          = {\url{https://andor.oxinst.com/assets/uploads/products/andor/documents/andor-marana-scmos-specifications.pdf}},
  note         = {Accessed: 2026-05-20}
}

@article{OHD_array1,
  title = {Quantum State Tomography with Array Detectors},
  author = {Beck, M.},
  journal = {Phys. Rev. Lett.},
  volume = {84},
  issue = {25},
  pages = {5748--5751},
  numpages = {0},
  year = {2000},
  month = {Jun},
  publisher = {American Physical Society},
  doi = {10.1103/PhysRevLett.84.5748},
  url = {https://link.aps.org/doi/10.1103/PhysRevLett.84.5748}
}

@article{OHD_array2,
  title = {Simultaneous quantum-state measurements using array detection},
  author = {Dawes, A. M. and Beck, M.},
  journal = {Phys. Rev. A},
  volume = {63},
  issue = {4},
  pages = {040101(R)},
  numpages = {4},
  year = {2001},
  month = {Mar},
  publisher = {American Physical Society},
  doi = {10.1103/PhysRevA.63.040101},
  url = {https://link.aps.org/doi/10.1103/PhysRevA.63.040101}
}

@article{PVLAS,
title = {The {PVLAS} experiment: A 25 year effort to measure vacuum magnetic birefringence},
journal = {Phys. Rep.},
volume = {871},
pages = {1-74},
year = {2020},
issn = {0370-1573},
doi = {https://doi.org/10.1016/j.physrep.2020.06.001},
url = {https://www.sciencedirect.com/science/article/pii/S0370157320302428},
author = {A. Ejlli and F. {Della Valle} and U. Gastaldi and G. Messineo and R. Pengo and G. Ruoso and G. Zavattini}
}

@Article{BMV,
author={Agil, J. and Battesti, R. and Rizzo, C.},
title={Monte Carlo study of the {BMV} vacuum linear magnetic birefringence experiment},
journal={Eur. Phys. J. D},
year={2021},
month={Mar},
day={08},
volume={75},
number={3},
pages={90},
issn={1434-6079},
doi={10.1140/epjd/s10053-021-00100-z},
url={https://doi.org/10.1140/epjd/s10053-021-00100-z}
}

\end{document}